\documentclass[symmetry,article,accept,moreauthors,pdftex]{Definitions/mdpi} 

\firstpage{1} 
\pubvolume{13}
\issuenum{193}
\articlenumber{193}
\pubyear{2021}
\copyrightyear{2021}
\externaleditor{{Academic Editor: Andrea Lavagno}}%MDPI: please check is there need Academic Editor?
\datereceived{10 December 2020} 
\dateaccepted{22 January 2021} 
\datepublished{26 January 2021} 
\hreflink{https://doi.org/10.3390/ sym13020193} % Please use \linebreak if need to do line break

\Title{An Analytical Approach to the Universal Wave Function and Its Gravitational Effect}

\TitleCitation{{An Analytical Approach to the Universal Wave Function and Its Gravitational Effect}}

\Author{{Yunuo Xiong}, Hongwei Xiong *} 
\AuthorNames{Yunuo Xiong and Hongwei Xiong}

\address[1] {%
{College of Science, Zhejiang University of Technology}, Hangzhou 31023, China; xyn20030409@zjut.edu.cn\\} %MDPI: the two affiliations are the same, so we delete one
\corres{\hangafter=1 \hangindent=1.05em \hspace{-0.82em}Correspondence: xionghw@zjut.edu.cn}

\abstract{Based on quantum origin of the universe, in this article we find that the universal wave function can be far richer than the superposition of many classical worlds studied by Everett. By analyzing the more general universal wave function and its unitary evolutions, we find that on small scale we can obtain Newton's law of universal gravity, while on the scale of galaxies we naturally derive gravitational effects corresponding to dark matter, without modifying any physical principles or hypothesizing the existence of new elementary particles. We find that an auxiliary function having formal symmetry is very useful to predict the evolution of the classical information in the universal wave function.}

\keyword{universal wave function; dark matter; gravitational force; symmetric formal solution}  % List three to ten pertinent keywords specific to the article, yet reasonably common within the subject discipline.

\begin{document}
%%%%%%%%%%%%%%%%%%%%%%%%%%%%%%%%%%%%%%%%%%

%%%%%%%%%%%%%%%%%%%%%%%%%%%%%%%%%%%%%%%%%%
\setcounter{section}{0} %% Remove this when starting to work on the template.

\section{Introduction}
The principal methodology of quantum physics is to find appropriate wave functions for complicated quantum systems and manifest their fundamental properties. Finding appropriate single-particle wave functions can greatly enrich our understanding of complicated systems, such as the eigenfunctions of infinite potential well and periodic potential for studies of crystal. Moreover, finding appropriate multi-body wave functions proved to be helpful for our understanding and design of complicated quantum material; for examples, the BCS wave function in superconductor \cite{BCS} and Laughlin wave function \cite{Laughlin} in fractional quantum hall effect. The most complicated wave function we can imagine ought to be the universal wave function. Everett analyzed formal solutions for the universal wave function \cite{Everett} and presented predictions which changed our understanding of the universe, such as the predictions that there are many classical worlds in the universe and those classical worlds are constantly branching. Unfortunately, the formal solution given by Everett does not have observational and falsifiable predictions and thus it failed to become a realistic physical theory at the present stage. 

In this article, we think that the universal wave function can be much richer than Everett's formal solution, by taking adequate consideration of quantum origin \cite{universe} of the universe. We discover that the diffusion particle wave packets due to quantum origin of the universe can have gravitational interactions with our classical world. We propose a general method to extract quantum information in the universal wave function; by using it we can recover Newton's law of universal gravity on small scale, while on the scale of galaxies we obtain the gravitational effects corresponding to dark matter, without modifying any physical principles or hypothesizing the existence of new elementary particles. We also discover the usual Newtonian gravitational constant can be different from the gravitational constant between elementary particles. 
Our work clearly shows that the universal wave function is not only realistic, but we can also obtain wave packets distribution information of the universal wave function through gravitational effects.

\section{The Unitary Evolutions of Universal Wave Function}
Mathematically, if we apply quantum mechanics to the entire universe, then at least under non-relativistic framework, we can assume the existence of a universal wave function $\Psi_u(\vec r_1,\cdots,\vec r_N,t)$ \cite{Everett}. Since we are working under non-relativistic framework, we do not consider those zero-mass particles, nor do we consider the creation and annihilation of particles. The universal wave function satisfies the linear unitary evolution of quantum mechanics:
\begin{equation}
i\hbar\frac{\partial\Psi_u}{\partial t}(\vec r_1,\cdots,\vec r_N,t)=(\hat H_0+\hat H_{esw}+\hat H_g)\Psi_u(\vec r_1,\cdots,\vec r_N,t),
\label{Schrodinger}
\end{equation}
where $\hat H_0$ denotes the kinetic energy operator of all the particles, $\hat H_{esw}$ denotes interactions except gravitational interactions. $\hat H_g$ denotes gravitational interactions. 
\begin{equation}
\hat H_g=-\frac{1}{2}G\sum_{l\neq n=1}^N\frac{m_l m_n}{|\vec r_l-\vec r_n|},
\label{Hg}
\end{equation}
where $G$ denotes the gravitational constant between elementary particles, $m_l$ is mass of the $l$th particle. Another equivalent way to consider the above Schr\"odinger equation is the Feynman path integrals for the universal wave function \cite{Bar}. 

Even if we ignore the internal freedom of elementary particles, the universal wave function will still be extremely complicated. Therefore we need to use suitable mathematical tools to analyze the universal wave function. Mathematically, we can define the single-particle wave packet distribution function as 
\begin{equation}
\rho_j(\vec r_j,t)=\int\prod_{i(\neq j)}d^3\vec r_i  \left|\Psi_{u}(\vec r_1,\cdots,\vec r_N,t)\right|^2. 
\end{equation}
$\rho_j(\vec r_j,t)$ reflects the wave packet distribution of the $j$th particle in the universe. Larger $\rho_j(\vec r_j,t)$ implies that more wave packets are distributed at that location. Naturally we can generalize the above definition to two-particle joint wave packet distribution function as 
\begin{equation}
\rho_{l,h}(\vec r_l,\vec r_h,t)=\int\prod_{i(\neq {l,h})}d^3\vec r_i\left|\Psi_u(\vec r_1,\cdots,\vec r_N,t)\right|^2. 
\end{equation}

One may also generalize the definition to general n-particle joint wave packet distribution function $\rho_{1,\cdots, n}(\vec r_1,\cdots, \vec r_n,t)$.

\section{Classical World Wave Function, Classical Information and the Universal Wave~ Function}
To analyze precisely universal wave functions containing the classical world, we must first understand the fundamental properties of wave functions behind the classical world. In our classical world most particles are formed by quantum correlation with nearby particles. Here quantum correlation means that the wave function of two or more particles cannot be written in a product form. For example, in hydrogen atom, $\psi(\vec r_p,\vec r_e,t)$ cannot be expressed as $\psi(\vec r_p,\vec r_e,t)=\phi_p(\vec r_p,t)\phi_e(\vec r_e,t)$, this implies there is quantum correlation between the electron and proton within a hydrogen atom; the same holds for the three quarks within a proton, which is why we regard hydrogen atoms and protons as composite particles. Given that the quantum correlation between the three quarks of a proton is extremely robust, we know the rest mass of every proton is the same and can be regarded as a fundamental physical constant.
\par
There should be a wave function to describe a given classical world in which particles are distributed in localized regions. We call such a wave function the classical world wave function $\Psi_c(\vec r_1,\cdots,\vec r_N,t)$, to distinguish it from the universal wave function $\Psi_u(\vec r_1,\cdots,\vec r_N,t)$. From our understanding of the classical world, the classical world wave function should satisfy the following necessary conditions.

\begin{enumerate}
\item The first necessary condition satisfied by $\Psi_c(\vec r_1,\cdots,\vec r_N,t)$ is: almost every particle in $\Psi_c(\vec r_1,\cdots,\vec r_N,t)$ has some sort of quantum correlation with one or more other particles. Such quantum correlations distribute over the entire universe in a network form in our classical world.

\item The second necessary condition for $\Psi_c(\vec r_1,\cdots,\vec r_N,t)$ goes like this: in our classical world, almost every particle is localized and distributed in a small region $\Sigma_j(t)$. 

For classical world, we require that the volume of $\Sigma_j(t)$ be on microscopic scale for most particles. 
Under this condition, the sets $\{\Sigma_j(t),j\in\{1,\cdots,N\}\}$ and $\{\vec r_j(t),j\in\{1,\cdots,N\}\}$ represent the classical information in the classical world within the universal wave function, here $\vec r_j(t)$ is about the center of region $\Sigma_j(t)$ and defines the classical location of a particle. The gravitational effects we observe in reality correspond to evolution of the set $\{\vec r_j(t),j\in\{1,\cdots,N\}\}$ over time.

\item Finally, the third condition for $\Psi_c(\vec r_1,\cdots,\vec r_N,t)$ is as follows: when $\vec r_l$ and $\vec r_n$ are separated by macroscopic distance, the quantum correlation between those two particles can be neglected. More precisely, we have the decomposition $\rho_{l,n}\simeq\rho_l\rho_n$.

\end{enumerate}

From the above three necessary conditions, we can see that the classical world wave function can only reflect partial information of the universal wave function, since in the localized particle regions $\Psi_c(\vec r_1,\cdots,\vec r_N,t)=\Psi_u(\vec r_1\in\Sigma_1(t),\cdots,\vec r_j\in\Sigma_j(t),\cdots,\vec r_N\in\Sigma_N(t),t)$, whereas in other regions $\Psi_c(\vec r_1,\cdots,\vec r_N,t)=0$ based on our request of the classical world. This shows clearly that $\Psi_c(\vec r_1,\cdots,\vec r_N,t)\neq \Psi_u(\vec r_1,\cdots,\vec r_N,t)$. It is clear that $\Psi_u$ is more fundamental than $\Psi_c$.

Generally speaking, the universe wave function is much more complex than the classical world wave function, considering the quantum origin \cite{universe} of our universe. Hence, we need a general way to get the classical world wave function and the corresponding classical information if we know the exact universal wave function.
Here we present a general method to search for classical worlds of a given universal wave function.

Choose two arbitrary particles $i$ and $j$, compute $\rho_i(\vec r_i,t)$ and $\rho_j(\vec r_j,t)$ from \linebreak $\Psi_u(\vec r_1,\cdots,\vec r_N,t)$, where we assume $\rho_i(\vec r_i,t)$ and $\rho_j(\vec r_j,t)$ each has $P(t)$ peaks. Now we can construct $P^2(t)$ different functions as follows:
\begin{equation}
\Psi_{(\Sigma_i^{\kappa}\Sigma_j^\iota)}(\vec r_1,\cdots,\vec r_N,t)=\left\{ \begin{aligned}&c_{\kappa\iota}(t)\Psi_u\left(\vec r_1,\cdots,\vec r_N,t\right)~~~~~~~~if~\vec r_i\in\Sigma_i^\kappa(t)~and~\vec r_j\in\Sigma_j^\iota(t)\\&0~~~~~~~~if~\vec r_i\notin\Sigma_i^\kappa(t)~or~\vec r_j\notin\Sigma_j^\iota(t)\\\end{aligned}\right.
\end{equation}

Here $c_{\kappa\iota}(t)$ are normalization constants. From the above equation we have constructed a set of functions with $P^2(t)$ elements: $\{\Psi_{(\Sigma_i^{\kappa}\Sigma_j^\iota)}(\vec r_1,\cdots,\vec r_N,t),\kappa\in\{1,\cdots,P(t)\}$,\linebreak $\iota\in\{1,\cdots,P(t)\}\}$.
\par
For a given $(\kappa,\iota)$, if $\Psi_{(\Sigma_i^{\kappa}\Sigma_j^\iota)}(\vec r_1,\cdots,\vec r_N,t)$ has about $N$ peaks in $\{\Sigma_j(t),j\in\{1,\cdots,N\}\}$, and satisfies the three necessary conditions for the existence of classical worlds when all $\vec r_j$ are confined to the domains $\Sigma_j(t)$, then it is possible that we have found a classical world. Additional conditions may be imposed to check if that is indeed a stable classical world wave function. 
After we get a classical world wave function, the classical information $\{\Sigma_1(t),\cdots,\Sigma_N(t)\}$ will be obtained simultaneously. 

Of course, the above extraction of the classical world wave function and the corresponding classical information based on two particles can be generalized to n particles if two particles are not enough to get the definite classical world wave function.

After the definition of the classical information based on the universal wave function, the question is how we determine the evolutions of the classical information.
At first sight, we would adopt the following ``type-I'' program.

\begin{enumerate}

\item {At $t_0$, we get the classical information of the classical world from the universal wave function $\Psi_u(t_0)$ with the above method to extract the classical information.}

\item {To predict the evolution of the classical information, we use Newtonian mechanics to calculate the evolution of the classical information $\{\vec r_j(t),j\in\{1,\cdots,N\}\}$.}
\end{enumerate}

{Since there is no external observer to interact with particles in the universe and record their classical information, we should always analyze the universal wave function from unitary evolutions. When the universal wave function is applied to the whole universe, we should not use the concept of instantaneous wave packet collapse. It is clear that to consider in the most accurate way the evolution of the classical information, we should first solve the unitary evolution of the universal wave function, and then search the classical world from the universal wave function at different times. Hence, the improved ``type-II'' program should be:}

\begin{enumerate}

\item  {At $t_0$, we get the classical information $\{\vec r_j(t_0),j\in\{1,\cdots,N\}\}$ of the universal wave function $\Psi_u(t_0)$ by searching the classical world.}

\item{From the solution of the Schr\"{o}dinger equation, we get the universal wave function $\Psi_u(t)$ at a later time.}

\item{From $\Psi_u(t)$, we get the classical information $\{\vec r_j(t),j\in\{1,\cdots,N\}\}$ by another search of the classical world.}

\item {Summarize the classical law from the relation between $\{\vec r_j(t_0),j\in\{1,\cdots,N\}\}$ and $\{\vec r_j(t),j\in\{1,\cdots,N\}\}$.}

\end{enumerate}

{Of course, the ``type-II'' program will provide more accurate evolution of the classical information, compared with the ``type-I'' program. However, at first sight, it seems that there should not be significant difference between the ``type-I'' program and ``type-II'' programs. Nevertheless, we will show that our universe provides a natural example to demonstrate the difference between the ``type-I'' and ``type-II'' programs.}

{Since we do not know what the universal wave function looks like and how it evolves, the ``type-II'' program is not as useful in practice without special mathematical technique. In the following, we will attempt to construct a function $\Psi_\chi(t)$, which we call the $\chi$ function, which unlike $\Psi_c(t)$, takes into account the evolutions of the universal wave function itself and provides better approximation to predict the evolution of the classical information we observe in reality, compared with evolutions of the classical world wave function by Schr\"{o}dinger equation alone. This $\Psi_\chi(t)$ function has the merit that it can naturally take into account the observational data to predict the details of the future evolution.}

\section{An Analytical Approach to the Universal Wave Function}

In the 1950s, Everett \cite{Everett} analyzed the following formal solution for the universal wave function:
\begin{equation}
\Psi_u(\vec r_1,\cdots,\vec r_N,t)=\sum_{\kappa=1}^{P(t)}\alpha_{\kappa}(t)\Psi_c^\kappa(\vec r_1,\cdots,\vec r_N,t).
\label{Everett}
\end{equation}

Here every $\Psi_c^\kappa(t)$ represents the classical world wave function and Equation (\ref{Everett}) expresses that there are $P(t)$ different classical worlds.
Normally, people think the superposition of classical worlds constitutes the whole universal wave function. However, we consider that to be a big mistake. 
The viable formal solutions of the universal wave function can be much richer than Equation (\ref{Everett}). We will introduce a general method to give more general universal wave function in due course.
%In the following we do not attempt to derive accurate formal solutions for the universal wave function; rather, we present mathematical tools to analyze the general universal wave function $\Psi_u(t)$.

\begin{figure}[H]

\widefigure
%\centering 
\includegraphics[width=0.8\textwidth]{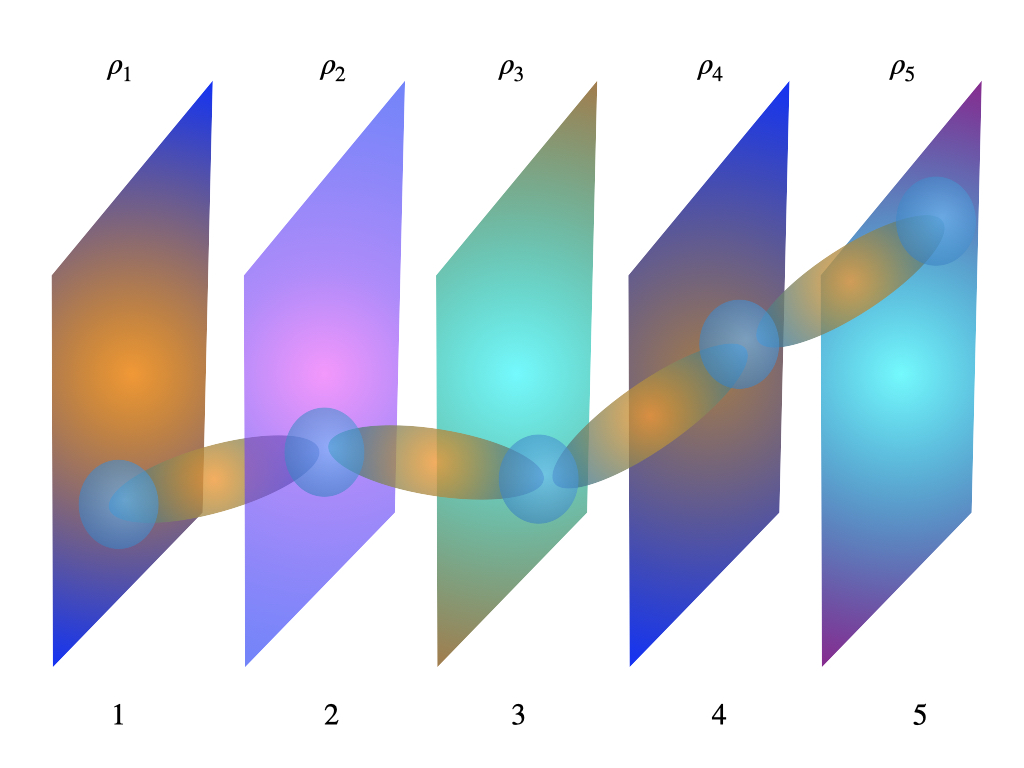} 
\vspace{-6pt}

\caption{As an example to illustrate the universal wave function and its classical world, we show the single-particle wave packet distribution for five particles. The quantum correlation between adjacent particles is also shown. We emphasize the possibility that there is a nonzero background distribution outside the region of the classical world; we can imagine this by thinking that at first the universal wave function is the product of the diffusion wave packets of those particles. Then we turn on interactions to form the classical world. It is clear that without an external observer, there must be a significant amount of diffusion wave packets in the universal wave function. The main result presented in this work is that those diffusion wave packets will influence the classical world through gravitational forces.}
\label{Fig1}
\end{figure}

For the formal solutions given by Equation (\ref{Everett}), it is sufficient to consider the evolutions of $\Psi_c(t)$ alone in Schr\"{o}dinger equation, and we will prove in Section \ref{acc} that the ``type-I'' and ``type-II'' programs will give the same evolution of the classical information. However, in the following we give a simple and realistic example to show that in general this is not the case.
For simplicity, we consider the case of a single classical world wave function. \mbox{Figure \ref{Fig1}} illustrates $\rho_j$ of the universal wave function for several particles. There is a common misunderstanding that $\Psi_u=0$ for any $\vec r_j\notin \Sigma_j(t)$. Everett's formal solution (\ref{Everett}) clearly satisfies this property. However, considering quantum origin of the universe, we cannot exclude the possibility that $\rho_j$ also has small background distribution. 
In this case, it is possible that $\Psi_u\neq 0$ for $\vec r_j\in\Sigma_j(t)$ and all other $\vec r_k\notin\Sigma_k(t)$. So in this case, we should consider to extend the domain of $\Psi_c$ to the entire universe to consider the long-range gravitational effect in a more accurate way.
\vspace{-12pt}

% start a new page without indent 4.6cm
%\clearpage
\end{paracol}
\nointerlineskip
\begin{paracol}{2}
%\linenumbers
\switchcolumn

As an example, we first consider the domain extension of the $j$th particle and define the function $\Psi_\chi^j(\vec r_1,\cdots,\vec r_N,t)$ so that $\vec r_l\in\Sigma_l(t)$ for all $l\neq j$, whereas there is no confinement on $\vec r_j$. In this case, we may rewrite the normalized $\Psi_\chi^j(t)$ as $\Psi_\chi^j(t)=\bar\Psi_c^j(t)\tilde\phi_j(\vec r_j,t)$.

\textls[-20]{Here $\bar\Psi_c^j(t)$ is defined as follows: if $\vec r_j$ is inside localized region $\Sigma_j(t)$, then $\bar\Psi_c^j(\vec r_1,\cdots,\vec r_N,t)$}\linebreak $=\Psi_c(\vec r_1,\cdots,\vec r_N,t)$, \textls[-10]{whereas if it is outside region} $\Sigma_j(t)$, then $\bar\Psi_c^j(\vec r_1,\cdots,\vec r_{j-1},\vec r_j,\cdots,\vec r_N,t)=\Psi_c(\vec r_1,...,\vec r_j(t),...,\vec r_N,t)/\sqrt{\rho_j(\vec r_j(t),t)}$. For $\vec r_j\notin \Sigma_j(t)$, note that $\vec r_j(t)$ denotes center of region $\Sigma_j(t)$ rather than the independent variable $\vec r_j$. For $\vec r_j\notin \Sigma_j(t)$, the number of 
independent variables in $\bar\Psi_c^j(\vec r_1,\cdots,\vec r_{j-1},\vec r_j,\cdots,\vec r_N,t)$ is $3(N-1)+1$. $\rho_j(\vec r_j(t),t)$ is the single-particle wave packet distribution function of $\Psi_c(t)$ at location $\vec r_j(t)$. 

\par
$\tilde\phi_j(\vec r_j,t)$ has the following properties: (i) when $|\vec r_j-\vec r_j(t)|<R_j$, $|\tilde\phi_j(\vec r_j,t)|^2=\gamma_j(t)$; (ii) when $|\vec r_j-\vec r_j(t)|>R_j$, $\tilde\phi_j(\vec r_j,t)=\sqrt{\zeta_j(t)}\phi_j(\vec r_j,t)$, where $\phi_j(\vec r_j,t)$ is normalized and diffuses across the entire universe. In the latter case, $|\phi_j(\vec r_j,t)|^2$ is very small, so we can freely choose $R_j$ on macroscopic scale, for example one meter, as long as it is greater than the radius of the localized wave packet distribution region $\Sigma_j(t)$. Moreover, since $|\phi_j(\vec r_j,t)|^2$ is very small and $R_j$ is of macroscopic scale, $|\phi_j(\vec r_j,t)|^2$ is not very sensitive to the exact location of $\vec r_j(t)$. The normalization condition requires that $\gamma_j(t)+\zeta_j(t)=1$.

\par

We may consider the domain extension of other particles in the same way. It is clear that there is no quantum correlation between $\phi_j(\vec r_j,t)$ for different particles. Proceeding in this way, we obtain the following form for $\Psi_\chi(t)$ by the domain extension of all $\vec r_j$ based on $\Psi_c(t)$:
\begin{equation}
\begin{aligned}
\Psi_\chi(\vec r_1,\cdots,\vec r_N,t)=\overline\Psi_c(t)\tilde\phi_1(\vec r_1,t)\tilde\phi_2(\vec r_2,t)...\tilde\phi_j(\vec r_j,t)...\tilde\phi_N(\vec r_N,t),
\end{aligned}
\label{extended}
\end{equation}
where all the $\tilde\phi_j(\vec r_j,t)$ are defined in the same way as above. For $\overline\Psi_c(t)$ in this case, however, we need to consider for each particle whether it is in its localized region $\Sigma_j(t)$. That is, we need to define $\overline\Psi_c(t)$ as follows: if all the particles are inside their localized regions $\Sigma_j(t)$, then $\overline\Psi_c(\vec r_1,\cdots,\vec r_N,t)=\Psi_c(\vec r_1,\cdots,\vec r_N,t)$; if a subset $A$ (say $A=\{1,2,...,n\}$) of particles are outside their localized regions $\Sigma_j(t)$, while all others are inside their localized regions, then $\overline\Psi_c(\vec r_{n+1},...,\vec r_N,t)=\Psi_c(\vec r_1(t),...,\vec r_n(t),\vec r_{n+1},...,\vec r_N,t)/\sqrt{\rho_{1,...,n}(\vec r_1(t),...,\vec r_n(t),t)}$, where $\rho_{1,...,n}(t)$ is the joint wave packet distribution function of $\Psi_c(t)$ for those $n$ particles. We can obtain similar formulas for each subset of particles and together they constitute complete definition of $\overline\Psi_c(t)$.

\par

It may seem that the $\chi$ function $\Psi_\chi(t)$ was transformed from  $\Psi_c(t)$ and $\Psi_u(t)$ by a rather artificial mathematical operation, in the following we elucidate why the $\chi$ function we introduce here has physical origin. In the early universe the wave packet of each particle completely spread out across the universe and there was no definite stable structure formed, no classical world existed at that point. Then as the universe expands, wave packets continue to spread and from the unitary evolutions of universe, we know those wave packets cannot ``collapse'' based on unitary evolution, which means that the diffusion wave packets always exist in the universe. So how did the classical world take its form? Because of the short and long range interactions between those particles, quantum correlation will stimulate structures to be formed among the diffusion wave packets and the network structure of quantum correlation between nearby particles represents the classical world. In general $\rho_j(\vec r_j,t)$ for each particle looks like a series of spikes with a diffusion background, each spike represents a distinct classical world. So as long as the diffusion background is not equal to 0, we can imagine the existence of $\Psi_\chi(t)$ by extending the domain of $\Psi_c(t)$ to the entire universe.

From Equation (\ref{extended}), we see that information about $\Psi_c(t)$ is contained within $\Psi_\chi(t)$ with no changes. However, $\Psi_\chi(t)$ contains more information through $\{\gamma_j(t),\zeta_j(t),\phi_j(\vec r_j,t),j\in\{1,\cdots,N\}\}$. In this work we consider a random model in which all $\zeta_j(t)$ are random numbers (noise) between $0$ and $1$, while $\{\phi_j(\vec r_j,t),j\in\{1,\cdots,N\}\}$ are a series of functions with no correlation with each other as time proceeds. Considering the complexity of the universal wave function and its quantum origin, and considering the rapid movement of any particle in the classical world relative to the cosmos background, at each time we need to reconstruct $\Psi_\chi$ from $\Psi_u$ and $\Psi_c$, which makes this random model reasonable. Of course, in this model, there is no unitary evolution relation between $\Psi_\chi$ at different times. Hence, there are only continuous evolutions of a single classical world (if we do not consider the classical world branching of $\Psi_c$ by itself), even when $\Psi_\chi\neq \Psi_c$. The correlation time of $\gamma_j(t)$ can be estimated by considering that when there is no overlapping between $\Sigma_l(t)$ and $\Sigma_l(t+\tau)$ ($l$ denotes a nearby particle), $\{\gamma_j(t),\phi_j(\vec r_j,t)\}$ and $\{\gamma_j(t+\tau),\phi_j(\vec r_j,t+\tau)\}$ may be assumed to be independent. Based on this assumption, it is estimated that $\tau\sim10^{-15}$ for a particle in the Milky Way. In this estimation, we assume that the velocity of the Milky Way relative to cosmos background is about $200$~km/s, while the size of $\Sigma_l(t)$ is about $10^{-10}$~m.

If at time $t$, $\Psi_\chi(t)=\Psi_u(t)$, then it is not the random model we consider. 
However, we believe that the universal wave function is much richer than this simple case so that $\Psi_\chi$ may be significantly different from $\Psi_c$ and $\Psi_u$ at any time. Of course, only observation will determine which model is correct, because it is impossible to solve the Schr\"odinger equation for the universal wave function to prove this. We will show in due course the observational evidence of our model.

Notice that, in our model $\Psi_\chi$  is not a real wave function at all; it only provides a mathematical method to consider the correlation between our classical world and other parts of the universal wave function, which enables us to study its gravitational effects later. It is nonsense to calculate the long-time unitary evolutions of $\Psi_\chi(t)$. Only the unitary evolution of $\Psi_u(t)$ is always correct, while the unitary evolution of $\Psi_c(t)$ may provide approximation to true evolutions of the classical information. We will show that $\Psi_\chi(t)$ gives better predictions to the gravitational acceleration of the particles at time $t$ in the classical world than those obtained from the unitary evolution of $\Psi_c(t)$.

\par
Here, we show two more properties of $\Psi_\chi(t)$. From the general expression of $\Psi_\chi(t)$ given by Equation (\ref{extended}), we have

(1) For normalized $\Psi_\chi(t)$, $\rho_j(\vec r_j,t)=\gamma_j(t)\rho_j^c(\vec r_j,t)+\zeta_j(t)\rho_j^d(\vec r_j,t)$ for the $j$th particle. Here the normalized $\rho_j^c(\vec r_j,t)$ denotes the localized part distributing over region $\Sigma_j(t)$, whereas $\rho_j^d(\vec r_j,t)$ is the diffusion part.

\par
(2) For the original $\Psi_c(t)$, when two particles are separated by macroscopic distance, $\rho_{l,m}(\vec r_l,\vec r_m,t)= \rho_l^c(\vec r_l,t)\rho_m^c(\vec r_m,t)$. For $\Psi_\chi(t)$, there can be no quantum correlation between diffusion wave packets of different particles, so we can assume that the decomposition formula still holds, namely
\begin{equation}
\rho_{l,n}(\vec r_l,\vec r_n,t)=(\gamma_l(t)\rho_l^c(\vec r_l,t)+\zeta_l(t)\rho_l^d(\vec r_l,t))(\gamma_n(t)\rho_n^c(\vec r_n,t)+\zeta_n(t)\rho_n^d(\vec r_n,t)).
\label{dec}
\end{equation}
%Notice that from the wave function (\ref{diffusion}), we cannot obtain such decompositions, instead we get $\rho_{l,n}=|\alpha|^2\rho_l^c\rho_n^c+|\beta|^2\rho_l^d\rho_n^d$.

Before considering gravitational interaction, we first think about whether the existence of $\Psi_\chi~(\neq \Psi_c)$ will influence evolutions of the classical information in $\Psi_c(t)$ through $\hat H_{esw}$. In the classical approximation, evolutions of the classical information of a particle is manifested by acceleration $\ddot r_j(t)$. The most useful and efficient way is to calculate $<\Psi_\chi | \hat H_{esw}|\Psi_\chi >$ first.  Since the diffusion wave packets diffuse very sparingly over the universe, they will not cause short range interactions such as weak and strong interactions, and because the diffusion wave packets as a whole is electrically neutral, they will not cause electromagnetic interactions either. That is, $<\Psi_\chi(t) | \hat H_{esw}|\Psi_\chi (t)>$ will not contain the information $\{\phi_j(\vec r_j,t),j\in\{1,\cdots,N\}\}$ in $\Psi_\chi(t)$. This leads to the result that the diffusion wave packets will not influence evolutions of the classical information in $\Psi_c(t)$ through $\hat H_{esw}$. We will show in the following section that gravitational force is different.

\section{Gravitational Acceleration of Particle Wave Packets in the Classical World}
\label{acc}

To obtain the acceleration of a particle by inter-particle interactions in the classical approximation, the most useful and efficient way is to calculate the interaction energy first. Quantum mechanics tells us that,  for a given state $\Psi(t)$, the total gravitational interaction energy is determined by the joint wave packet distribution functions as follows:
\begin{equation}
E_g(t)=\left<\Psi(t)|\hat H_g|\Psi(t)\right>=-\frac{1}{2}G\sum_{l\neq n}m_l m_n\int d^3\vec r_l d^3\vec r_n\frac{\rho_{l,n}(\vec r_l,\vec r_n,t)}{|\vec r_l-\vec r_n|}.
\label{AEg}
\end{equation}

From the above formula, it is straightforward to obtain gravitational effects in the classical approximation. Here $\rho_{l,n}(\vec r_l,\vec r_n,t)$ is the two-particle joint wave packet distribution function of $\Psi(t)$, while $\hat H_g$ is given by Equation (\ref{Hg}).

If we calculate $E_g(t)$ of $\Psi_\chi(t)$, it is clear that $E_g(t)$ will contain all the information $\{\gamma_j(t),\zeta_j(t),\phi_j(\vec r_j,t),j\in\{1,\cdots,N\}\}$ in $\Psi_\chi(t)$, because the gravitational interaction is universal between any two particles. Hence, it is expected that gravitational acceleration $\ddot r_j(t)$ in the classical world will be influenced by the diffusion wave packets contained in $\Psi_\chi(t)$.
We will use the usual method to calculate the gravitational acceleration. For the wave packet of the $j$th particle in $\Sigma_j(t)$, we consider an infinitesimal displacement $\Delta \vec r_j$ of this wave packet with a group acceleration $\vec a_j(t)$ during an infinitesimal time interval $\Delta t$. Based on the fact that the increase of the average kinetic energy is the decrease of the gravitational interaction energy, we can get the expression of the group acceleration  $\vec a_j(t)$.

First we consider the simplest case: suppose the universal wave function consists of a single classical world and there are no diffusion parts, i.e., $\Psi_u(t)=\Psi_c(t)=\Psi_\chi(t)$. In this situation, when two particles are separated by macroscopic distance, we have
\begin{equation}
\rho_{l,n}(\vec r_l,\vec r_n,t)=\rho_l(\vec r_l,t)\rho_n(\vec r_n,t).
\end{equation}

Substituting into Equation (\ref{AEg}), we get
\begin{equation}
E_g(t)\approx -\frac{1}{2}\sum_{i\neq j}G\frac{m_im_j}{|\vec r_i(t)-\vec r_j(t)|}.
\label{Egsimple}
\end{equation}

Now we analyze the gravitational force for the $j$th particle. We assume that during an infinitesimal  time interval $\Delta t$ all of the momentum changes for this particle are caused by gravity. Assuming that the wave packet of this particle moved from $\vec r_j(t)$ to $\vec r_j(t+\Delta t)$, while its group velocity changed from $\vec v_j(t)$ to $\vec v_j(t+\Delta t)$. Then, based on energy conservation, we have
\begin{equation}
\frac{1}{2}m_j\left(\left|\vec v_j(t+\Delta t)\right|^2-\left|\vec v_j(t)\right|^2\right)=E_g(\vec r_j(t),\cdots)-E_g(\vec r_j(t+\Delta t),\cdots).
\end{equation}

First we consider the $x$-components $v_{j}^x$ and $r_j^x$, the left hand side of the above equation becomes:
\begin{equation}
\frac{1}{2}m_j\left((v_j^x(t+\Delta t))^2-(v_j^x(t))^2\right)= m_j v_j^x\frac{dv_j^x}{dt}\Delta t.
\end{equation}

Whereas the right hand side is:
\begin{equation}
E_g(\vec r_j(t),\cdots)-E_g(\vec r_j(t+\Delta t),\cdots)=-\frac{\partial E_g}{\partial r_j^x}\Delta r_j^x,
\end{equation}
where $\Delta r_j^x=r_j^x(t+\Delta t)-r_j^x(t)$. Since $v_j^x=\Delta r_j^x/\Delta t$, we have the group acceleration
\begin{equation}
a_j^x=\frac{dv_j^x}{dt}=-\frac{1}{m_j}\frac{\partial E_g}{\partial r_j^x}.
\end{equation}

Using Equation (\ref{Egsimple}), we obtain the following acceleration formula in the classical approximation:
\begin{equation}
\begin{aligned}\vec a_j=&\sum_{i(\neq j)}G\frac{m_i}{|\vec r_i(t)-\vec r_j(t)|^3}(\vec r_i(t)-\vec r_j(t)).\end{aligned}
\end{equation}

This is the equation of Newton's law of universal gravity.

Now we analyze the formal solution (\ref{Everett}) given by Everett. In this formal solution, every $\Psi_c^\kappa$ represents a classical world and there are still no diffusion terms, i.e., if we perform the domain extension for any $\Psi_c^\kappa(t)$, we have $\Psi_c^\kappa(t)=\Psi_\chi^\kappa(t)$. The two-particle joint wave packet distribution function of $\Psi_u(t)$ becomes:
\begin{equation}
\rho_{l,n}(\vec r_l,\vec r_n,t)=\sum_{\kappa=1}^{P(t)}|\alpha_\kappa(t)|^2\rho_l^\kappa(\vec r_l,t)\rho_n^\kappa(\vec r_n,t),
\end{equation}
where normalized $\rho_l^\kappa(\vec r_l,t)$ represents the single-particle wave packet distribution function for the $l$th particle in the $\kappa$th classical world within region $\Sigma_j^\kappa(t)$. Thus, the total gravitational interaction energy of $\Psi_u(t)$ is
\begin{equation}
E_g(t)\approx -\frac{1}{2}\sum_{\kappa=1}^{P(t)}|\alpha_\kappa(t)|^2\left(\sum_{i\neq j}G\frac{m_im_j}{|\vec r_i^\kappa(t)-\vec r_j^\kappa(t)|}\right).
\label{EgEverett}
\end{equation}

Note that there are no gravitational interactions between different classical worlds in this case, because there is no gravitational interaction energy between different classical worlds.
\par
For this universal wave function, we can obtain the gravitational acceleration formula for the $j$th particle in the $\kappa$th classical world from the following relation:
\begin{equation}
\frac{1}{2}|\alpha_\kappa(t)|^2m_j\left(\left|\vec v_j^\kappa(t+\Delta t)\right|^2-\left|\vec v_j^\kappa(t)\right|^2\right)=E_g(\vec r_j^\kappa(t),\cdots)-E_g(\vec r_j^\kappa(t+\Delta t),\cdots).
\end{equation}
$|\alpha_\kappa(t)|^2$ on the left and right hand sides cancel out, so we have
\begin{equation}
\begin{aligned}
\vec a_j^\kappa=&\sum_{i(\neq j)}G\frac{m_i}{|\vec r_i^\kappa(t)-\vec r_j^\kappa(t)|^3}(\vec r_i^\kappa(t)-\vec r_j^\kappa(t)).
\label{acc17}
\end{aligned}
\end{equation}

This implies each classical world evolves independently according to Newton's law of universal gravity and such evolutions do not depend on $|\alpha_\kappa(t)|^2$. Based on  Everett's universal wave function (\ref{Everett}), at macroscopic scale, we can not notice different gravitational effect, compared with Newton's universal law of gravity.
\par

For the following uncorrelated diffusion universal wave function,
\begin{equation}
\Psi_u(t)=\sum_{\kappa=1}^{P(t)}\alpha_{\kappa}(t)\Psi_c^\kappa(\vec r_1,\cdots,\vec r_N,t)+\beta(t)\prod_{j=1}^N\phi_j^d(\vec r_j,t),
\label{AEverett}
\end{equation}
similar calculations show that the gravitational accelerations are still given by \mbox{Equation (\ref{acc17})}, since the gravitational interaction between each classical world and $\beta(t)\prod_{j=1}^N\phi_j^d(\vec r_j,t)$ is 0. In the above expression, we assume that all $\phi_j^d$ are diffused in the entire universe.
\par

In Everett's universal wave function given by Equation (\ref{Everett}), for each $\Psi_c^\kappa(\vec r_1,\cdots,\vec r_N,t)$, the domain is highly localized. We have emphasized previously that this is a very specialized universal wave function. When diffusion wave packets are fully taken into account, we should use $\Psi_\chi(t)$ to calculate the gravitational acceleration at $t$.
For a given $\Psi_\chi(t)$, based on the formula given in this article, for two particles separated by macroscopic distance, we have the two particle joint wave packet distribution function given by Equation (\ref{dec}).
%\begin{equation}
%\rho_{ln}(\vec r_l,\vec r_n,t)=\rho_l(\vec r_l,t)\rho_n(\vec r_n,t)=(\gamma_l(t)\rho_l(\vec r_l,t)+\zeta_l(t)\rho_l(\vec r_l,t))(\gamma_n(t)\rho_n(\vec r_n,t)+\zeta_n(t)\rho_n(\vec r_n,t)).
%\label{chiexpan}
%\end{equation}

For the case of a single classical world, we may always decompose $\Psi_u(t)$ as 
\begin{equation}
\Psi_u(t)=\alpha_\chi(t)\Psi_\chi(t)+\beta(t)\Psi_r(t).
\label{decomp}
\end{equation}
%Here
%\begin{equation}
%\Psi_r(t)=\prod_{j=1}^N\phi_j^r(\vec r_j,t)
%\end{equation}
%with every $\phi_j^r(\vec r_j,t)$ diffused in the entire universe.

Here the term $\beta(t)\Psi_r(t)$ is due to the general consideration that $\Psi_u(t)\neq \Psi_\chi(t)$. For any $\Psi_u$ and $\Psi_{\chi}$, we can always make this decomposition. We may assume further that 
 \begin{equation}
 <\Psi_\chi(t)|\Psi_r(t)>=0.
 \end{equation}
 
It is clear that the superposition principle of quantum mechanics can always make the appropriate decomposition (\ref{decomp}) to satisfy this request.
 
The total gravitational interaction energy of $\Psi_u(t)$ is therefore:
\begin{equation}
E_g(t)\approx |\alpha_\chi(t)|^2 E_g^\chi(t)+|\beta(t)|^2E_g^r(t).
\end{equation}

Here $E_g^\chi(t)$ is the total gravitational interaction energy of $\Psi_\chi(t)$, while $E_g^r(t)$ is the total gravitational interaction energy of $\Psi_r(t)$. It is clear that the term $<\Psi_\chi(t)|\hat H_g|\Psi_r(t)>$ is negligible. The above expression of $E_g(t)$ is the motivation of the introduction of the $\chi$ function $\Psi_\chi$ to calculate the gravitational effect in the universal wave function. From Equations (\ref{dec}) and (\ref{AEg}), we get
\begin{equation}
\begin{aligned}
E_g^\chi(t)\approx &-\frac{1}{2}\sum_{i\neq j}G \gamma_i(t)\gamma_j(t)\frac{m_im_j}{|\vec r_i(t)-\vec r_j(t)|}\\&-\sum_{i\neq j}G \gamma_i(t)\zeta_j(t)\int d^3\vec r_j\frac{m_i m_j\rho_j^d(\vec r_j,t)}{|\vec r_i(t)-\vec r_j|}\\&-\frac{1}{2}\sum_{i\neq j}G\zeta_i(t)\zeta_j(t)\int d^3\vec r_i\int d^3\vec r_j\frac{m_i\rho_i^d(\vec r_i,t) m_j \rho_j^d(\vec r_j,t)}{|\vec r_i-\vec r_j |},
\label{Appenergy}
\end{aligned}
\end{equation}
where $\vec r_j(t)$ denotes the classical position of the $j$th particle within region $\Sigma_j(t)$, whereas $\vec r_j$ denotes position variable to be integrated.
\par
Now we consider the gravitational force for the wave packet of the $j$th particle in the classical world, i.e., in the region $\Sigma_j(t)$. Again, we assume during infinitesimal time interval $\Delta t$, all momentum changes for this wave packet are caused by gravity, we have
\begin{equation}
\gamma_j(t)\frac{1}{2}m_j\left(\left|\vec v_j(t+\Delta t)\right|^2-\left|\vec v_j(t)\right|^2\right)=E_g^\chi(\vec r_j(t),\cdots)-E_g^\chi(\vec r_j(t+\Delta t),\cdots).
\end{equation}

For $v_{j}^x$ and $r_{j}^x$, we have
\begin{equation}
\gamma_j(t)\frac{1}{2}m_j\left((v_j^x(t+\Delta t))^2-(v_j^x(t))^2\right)=\gamma_j(t) m_j v_j^x\frac{dv_j^x}{dt}\Delta t,
\end{equation}
and
\begin{equation}
E_g^\chi(\vec r_j(t),\cdots)-E_g^\chi(\vec r_j(t+\Delta t),\cdots)=-\frac{\partial E_g^\chi}{\partial r_j^x}\Delta r_j^x.
\end{equation}

In this case, we have
\begin{equation}
a_j^x=\frac{dv_j^x}{dt}=-\frac{1}{\gamma_j(t) m_j}\frac{\partial E_g^\chi}{\partial r_j^x}.
\end{equation}

Therefore,
\begin{equation}
\begin{aligned}
\vec a_j=&\sum_{i(\neq j)}G\frac{\gamma_i(t)m_i}{|\vec r_i(t)-\vec r_j(t)|^3}(\vec r_i(t)-\vec r_j(t))+\sum_{i(\neq j)}G\int d^3\vec r_i\frac{ \zeta_i(t)m_i\rho_i^d(\vec r_i,t)}{|\vec r_i-\vec r_j(t)|^3}(\vec r_i-\vec r_j(t)).\\
\end{aligned}
\end{equation}

It is worth pointing out that the value of $\gamma_j(t)$ does not appear in the gravitational acceleration $\vec a_j(t)$ of the $j$th particle. Hence, the fluctuation of the gravitational acceleration of a particle is due to the fluctuations of $\gamma_l(t)$ of other particles.

We assume that the probability distribution of the value of $\gamma_i(t)$ at different times is the same for every particle. Hence, we may use the time average values $\gamma$ and $\zeta$ to replace $\gamma_i(t)$ and $\zeta_i(t)$ in the above equation. $\gamma$ and $\zeta$ can also be regarded as the averages for relevant particles.
When gravitational effects due to the fluctuations of $\gamma_i(t)$ and $\zeta_i(t)$ are omitted, we have
\begin{equation}
\begin{aligned}
\vec a_j=&\sum_{i(\neq j)}G_N\frac{m_i}{|\vec r_i(t)-\vec r_j(t)|^3}(\vec r_i(t)-\vec r_j(t))+\sum_{i(\neq j)}G_N\int d^3\vec r_i\frac{ dm_i\rho_i^d(\vec r_i,t)}{|\vec r_i-\vec r_j(t)|^3}(\vec r_i-\vec r_j(t))\\
\label{acceleration}
\end{aligned}
\end{equation}
where the second term denotes the gravitational acceleration due to diffusion wave packets in $\Psi_\chi(t)$. 
 %In addition, we can not exclude the possibility that different types of fundamental particles may have different distribution of $\gamma_j$.
 In the above equation, 
 \begin{equation}
 G_N=G\gamma=\frac{G}{(d+1)}.
 \end{equation}
 
The parameter $d=\zeta/\gamma$ characterizes the gravitational interaction due to the diffusion wave packets in $\Psi_\chi(t)$.

\par
For particles in the classical world, from a gravitational viewpoint, the second term in Equation (\ref{acceleration}) suggests that the equivalent mass of the diffusion part is thus $M_d=dM_u$, where $M_u=\sum_j m_j$ is the sum of the actual masses of all the $N$ particles in the universe. Since the only way in which the diffusion wave packets can affect the classical world is through gravitational interactions, we call $M_d$ "fictitious" gravitational matter.
\par
When we observe gravitational effects in a small region, the contribution from diffusion part can be neglected and the gravitational constant we obtain through such observation is $G_N$, which we call Newtonian gravitational constant. On the other hand, the constant $G$ is a constant related to elementary particles and is different from $G_N$, we call $G$ the gravitational constant of elementary particles. After we obtain $G_N$, we can derive $G$ through $G=(d+1)G_N$. 

\section{Applications to the Gravitational Effect of Dark Matter}
The mathematical model of $\Psi_\chi$ provides a way to determine the size of $d$. Therefore, $d$ becomes a parameter originating from many-body evolutions of quantum mechanics, like lattice constant of a crystal. We cannot derive $d$ from first principles, but we can experimentally measure the size of $d$.
\par
The method for measuring $d$ is as follows: first based on Equation (\ref{acceleration}), we can omit the second term and measure the value of $G_N$ from earth-based small macroscopic scale experiments. Next we analyze the motion of objects on a larger scale, such as the motion of objects on the edge of galaxy. Therefore we can obtain $d$ and $M_d$ from the difference between the observed $\vec a_j$ and the theoretical result of the first term in Equation (\ref{acceleration}). Fortunately, the remarkable advances on the astronomical observations make this scheme completely feasible. From the value of $G_N$ and $d$ by observations, we get the true gravitational constant $G$ for elementary particles.

As we all know, gravitational effects for objects on the edge of galaxy are not determined by the first term in Equation (\ref{acceleration}) alone \cite{Dark}. In fact, $\vec a_j$ calculated from the first term in Equation (\ref{acceleration}) is several times smaller than the observed $\vec a_j$. Up to now, people still could not explain this difference and they hypothesized so called dark matter. Now we naturally discover the possibility that, the ``gravitational matter'' discussed here can be fit into the dark matter model.
\par

If we define the classical world mass density distribution function by 
\begin{equation}
\rho_m(\vec r,t)=\sum_jm_j\rho_j^c(\vec r,t)
\end{equation}
and gravitational matter mass density distribution function by 
\begin{equation}
\rho_d(\vec r,t)=d\sum_jm_j\rho_j^d(\vec r,t), 
\end{equation}
then based on all the observational and theoretical researches of dark matter \cite{Dark}, we can reasonably hypothesize that $\rho_d(\vec r,t)$ may have large scale structure. In addition, it seems that we may regard $\rho_d(\vec r,t)/d$ as the mass density distribution of $\Psi_u(t)$. {This means that from the observed ``dark matter'' distribution $\rho_d(\vec r,t)$ and the numerical simulation of the evolution of the universal wave function, we have the chance to know the initial wave function of the universe, which plays a special role in the initial state \cite{Gorobey} of inflation theory \cite{Guth}. Of course, to give an accurate deduction of the initial condition at the inflation stage, we need to improve our theory to the relativistic version of quantum gravity, such as the possible application of the Wheeler--DeWitt equation \cite{Wheeler}. It is worth pointing out that the universe wave function considered in this paper is different from that of the wave function of the universe in \cite{Hartle}, where the formal solution of the Wheeler--DeWitt equation is considered to reveal quantum gravity. In our work, however, we do not consider the quantization of gravity; what we consider is the gravitational effect due to the diffusion wave packets after the quantum origin of our universe. Hence, in Equation (\ref{acceleration}), the right hand side does not contain a term comprising $\hbar$ term, while we expect that further improvement by including the quantization of gravity will lead to a correction of $\hbar$ term.}
%这就提供了手段去研究暴涨并检验我们的理论。
\par
From the present observational results, we have $d\approx 5.5$ \cite{Dark}, this means the diffusion part can be rather significant.
Under non-relativistic framework, the total particle number $N$ of the universe is conserved. Thus in a given classical world, no matter how we count the number of total particles with a detector, the detectable total particle number is always $N$. Therefore, for observers in the classical world, the total mass of the universe they obtain by counting the particle number is always $M=\sum_j m_j$. Hence, the observer should not think that the gravitational matter are some extra particles, although the observer will notice its gravitational effect. It is obvious that our conclusions still hold when we consider the creation and annihilation of particles in the classical world. Because we use non-relativistic Schr\"{o}dinger equation in this work to consider the classical approximation of gravitational effect, the application of Equation (\ref{acceleration}) should be confined to the scale of galaxies, which is sufficient for the present application. For larger scale of the universe, we should improve our model to consider the curved spacetime of general relativity.
\par
The fact that there are no unitary relations between $\Psi_\chi$ at different times is not surprising. If we consider classical world branching, there are no unitary relations between $\Psi_c$ at different times even if there are no diffusion wave packets. Suppose we do not consider diffusion wave packets, then the only requirement for the existence of classical world at a given time $t$ is $<\Psi_c(t)|\hat U(t,t_0)|\Psi_u(t_0)>\neq 0$, here $\Psi_u(t_0)$ is the universal wave function at an initial time $t_0$. For $\Psi_\chi(t)$ this is similar, it is hard for us to answer why at time $t$ we can construct such $\Psi_\chi(t)$ from $\Psi_c(t)$ and $\Psi_u(t)$ and how to get the exact $\Psi_\chi(t)$, since we do not know the exact form of $\Psi_u(t)$ at time $t$ in the first place. However, so long as $<\Psi_\chi(t)|\hat U(t,t_0)|\Psi_u(t_0)>\neq 0$ is satisfied, in principle we can regard $\Psi_\chi(t)$ as a function of some significance. These analysis show that the sensible question to ask for future studies is what kind of initial universal wave function can cause $\Psi_\chi(t)$ to be unorthogonal to $\Psi_u(t)$, where $\Psi_\chi(t)$ contains $\Psi_c(t)$.
\par
In the formal solution given by Everett \cite{Everett}, there are no longer any diffusion wave packets after some phase transitions occurred in the early universe in which particle wave packets are completely spread out. From the point of view of unitary evolutions, there is no such so called wave packet collapse, so the more natural hypothesis is that after phase transitions occurred there are mixtures of diffusion parts and correlated localized distribution of wave packets. The parameter $d$ reflects the ratio of such mixtures. This inspires us that the process by which the universe formed its structures is a kind of phase transition in which diffusion and localized distributions coexist. After the phase transition takes place, during long periods of evolution of the universe the information about $d$ is retained and it has two general characteristics: (i) since different classical worlds are formed from the same universal wave function, we expect that the values of $d$ of $\Psi_\chi$ corresponding to other classical worlds should be about the same as our world; (ii) suppose at a time $t_1$ we have $d(t_1)\neq 0$, then we think at a later time $t_2$ we have $d(t_2)=d(t_1)$. Of course, this is not a proof, but an assumption of the universal value of the parameter $d$. It is similar to the formation of crystal that, the lattice constant of the crystal is always the same, although the initial condition to form the crystal can be significantly different, and we can not prove it by a calculation of the real evolutions.
\par

\section{An Improved Schr\"{o}dinger Equation for Classical World Wave Function}

The central purpose of the present work is to calculate evolutions of the classical information $\{\Sigma_1(t),\cdots,\Sigma_N(t)\}$ from unitary evolutions of $\Psi_u(t)$, because only the classical information can be recorded by observers. The ideal way for this is to first solve the evolution of $\Psi_u(t)$, and search for the classical world at different times from $\Psi_u(t)$. From the classical world wave function $\Psi_c(t)$ obtained from $\Psi_u(t)$, we can know the evolution of the classical information $\{\Sigma_1(t),\cdots,\Sigma_N(t)\}$. In the case of $\Psi_c=\Psi_\chi=\Psi_u$, we may get the classical approximation of the gravitational acceleration without using the method in this work. In the case of $\Psi_c\neq\Psi_\chi$, $\Psi_\chi\neq\Psi_u$ and $\Psi_\chi(t_1)\neq \hat U(t_1,t)\Psi_\chi(t)$, however, special method should be used to consider the evolution of classical information $\{\Sigma_1(t),\cdots,\Sigma_N(t)\}$. 
Among these three kinds of wave functions, only the universal wave function satisfies unitary evolutions of quantum mechanics.
$\Psi_c$ and $\Psi_\chi$ are only functions derived from $\Psi_u$. The final result we get should focus on $\Psi_c$ and its corresponding classical information. It is nonsense to calculate the long time unitary evolution of $\Psi_\chi$.

\par
In Figure \ref{Fig2} we summarize relationships between these three kinds of functions. There are unitary evolutions between the universal wave function at different times, we indicated this in the top of the figure for the set of universal wave functions $\{\Psi_u(t_0),\Psi_u(t_1),\Psi_u(t_2)\}$ at times $\{t_0,t_1,t_2\}$. From the universal wave function we can always search for classical worlds, here $\{C_\kappa(t_0),C_\kappa(t_1),C_\kappa(t_2)\}$ represents the set of classical information related by temporal causal relationships, while $\{\Psi_c^\kappa(t_0),\Psi_c^\kappa(t_1),\Psi_c^\kappa(t_2)\}$ represents the set of classical world wave functions corresponding to $\{C_\kappa(t_0),C_\kappa(t_1),C_\kappa(t_2)\}$. The so called classical theorem (here denoted by $\hat C$) represents physical laws to predict how to evolve from $C_\kappa(t_0)$ to $C_\kappa(t_1)$ and then to $C_\kappa(t_2)$. Our analysis shows that, if we consider gravitational effects, then in general we cannot deduce good classical theorem from the unitary evolution of $\Psi_c^\kappa$, since the best way is to derive $\{C_\kappa(t_0),C_\kappa(t_1),C_\kappa(t_2)\}$ directly from the set of universal wave functions $\{\Psi_u(t_0),\Psi_u(t_1),\Psi_u(t_2)\}$. However, that is not a realistic approach so we introduced the set of functions $\{\Psi_\chi(t_0),\Psi_\chi(t_1),\Psi_\chi(t_2)\}$ to improve the classical theorem by considering more carefully the gravitational interaction, compared with the classical theorem obtained from the unitary evolution of $\Psi_c^\kappa$.

\begin{figure}[H]
%\centering
 \includegraphics[width=0.7\textwidth]{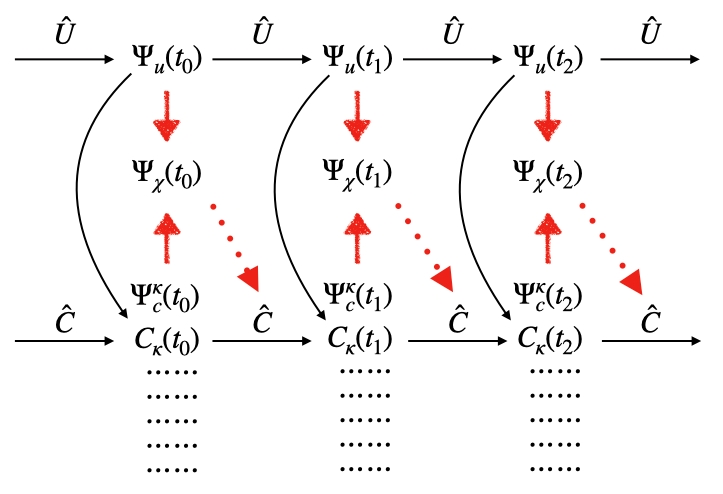} 
\caption{The relation between $\Psi_u,\Psi_c,\Psi_\chi$ and classical information.}
\label{Fig2}
\end{figure}

\par
Observe that if we do not consider gravity, the solution of $\Psi_c^\kappa$ from Schr\"{o}dinger equation is a natural mathematical method to describe evolutions of the classical world, we only need to take proper account of classical world branching. Thus, we naturally consider the question of whether corrections from the $\chi$ function would result in an improved Schr\"{o}dinger equation. We define a new function $\Psi_o^\kappa$ and let $\Psi_o^\kappa(t_0)=\Psi_c^\kappa(t_0)$. Then we think the following modified Schr\"{o}dinger equation for $\Psi_o^\kappa(t)$ reflects evolutions of the classical world in a reasonably accurate manner:

\begin{equation}
i\hbar\frac{\partial\Psi_o^\kappa}{\partial t}=\left(\hat H_0+\hat H_{esw}+\hat H_{G_N}-G_N\sum_j m_j\int d^3\vec x\frac{\rho_d(\vec x,t)}{|\vec r_j-\vec x|}\right)\Psi_o^\kappa.
\label{ESchrodinger}
\end{equation}

Here
\begin{equation}
\hat H_{G_N}=-\frac{1}{2}G_N\sum_{l\neq n=1}^N\frac{m_l m_n}{|\vec r_l-\vec r_n|},
\label{HGN}
\end{equation}
while the last term on the right hand side of Equation (\ref{ESchrodinger}) should be determined by astronomical observations.

It is easy to show that the classical theorem derived from the above Schr\"{o}dinger equation is the same as the classical theorem derived from combined consideration of $\Psi_c$ and $\Psi_\chi$. We call $\Psi_o^\kappa$ gravity-rectified classical world wave function. $G_N$ can also be regarded as a gravitational constant resulting from rectifying of gravitational constant of elementary particles. In the above equivalent Schr\"{o}dinger equation, we do not have to restrict domains of variables $\vec r_j$ in $\Psi_o^\kappa$. The merit of the above improved Schr\"odinger equation lies in that almost all the physical phenomenon are fully considered, such as the classical world branching, all gravitational effects including the gravitational force due to diffusion wave packets, and quantum process in our 
laboratory. In a sense, we may regard this equation as the final result of our work.

So far we have introduced four kinds of functions: the universal wave function $\Psi_u(t)$, the classical world wave function $\Psi_c(t)$, gravity-rectified classical world wave function $\Psi_o(t)$ and $\chi$ function $\Psi_\chi(t)$. Among them only the universal wave function is ultimate realistic entity and satisfies unitary evolutions of quantum mechanics. $\Psi_c(t)$ is also realistic in some sense, since we can learn classical information of each particle in the classical world from $\Psi_c(t)$. In that sense $\Psi_o(t)$ is similar to $\Psi_c(t)$, the difference being that when we solve Schr\"{o}dinger equation restricted in the classical world domain $\cup\Sigma_j(t)$, $\Psi_o(t)$ gives better prediction of the evolution of the classical information than $\Psi_c(t)$. From the point of view of perturbations, this is because $\Psi_o(t)$ takes better account of influences the universal wave function has on evolutions of the classical world. Without gravitational interaction, $\Psi_o(t)$ and $\Psi_c(t)$ are the same, and it is not necessary to consider $\Psi_\chi(t)$ even when we still have $\Psi_\chi\neq \Psi_c$ and $\Psi_\chi\neq \Psi_u$. As for $\Psi_\chi(t)$, in some sense it is just a function without any realistic value, which is why we do not call it wave function; it is just a helper function to deal with gravitational effects using perturbation method and the diffusion part in $\Psi_\chi(t)$ can change almost instantly. Hence, the classical world can not have memories of the information of $\{\zeta_j(t),\phi_j(\vec r_j,t)\}$ for each particle.

\par
Based on the above analysis, we see that evolutions of $\rho_d(\vec r,t)$ in Equation (\ref{ESchrodinger}) may not be calculated from unitary evolutions of $\Psi_\chi(t)$. $\rho_d(\vec r,t)$ may only be calculated from evolutions of the universal wave function.  In practice, we need to observe the distribution of $\rho_d(\vec r,t)$ through gravitational effects. Considering that $\Psi_\chi(t)$ needs to be reconstructed at every instant, what we observe is actually the temporal and spatial average of $\rho_d(\vec r,t)$, namely
\begin{equation}
\bar\rho_d(\vec r,t)=\frac{\int_{t-\Delta t/2}^{t+\Delta t/2}dt\int_{\Omega} d^3\vec y\rho_d(\vec y,t)}{V_\Omega\Delta t},
\end{equation}
where $\Omega$ in $\int_\Omega$ denotes the sphere with center $\vec r$, $V_\Omega$ denotes volume of the sphere.
\par
Thus the form of Equation (\ref{ESchrodinger}) can be improved as:
\begin{equation}
\left\{ \begin{aligned}&i\hbar\frac{\partial\Psi_o^\kappa(t)}{\partial t}=\left(\hat H_0+\hat H_{esw}+\hat H_{G_N}-G_N\sum_j m_j\int d^3\vec x\frac{\bar\rho_d(\vec x,t)}{|\vec r_j-\vec x|}\right)\Psi_o^\kappa(t)\\&\Psi_u(t)=\hat U(t,t_0)\Psi_u(t_0)\\&\{\Psi_o^\kappa(t),\Psi_u(t)\}\rightarrow \bar\rho_d(\vec x,t)\\\end{aligned}\right.
\label{simulation}
\end{equation}

In the above equation the evolution of $\bar\rho_d(\vec x,t)$ over time is not determined by unitary evolutions of $\Psi_\chi(t)$ (we have explained that $\Psi_\chi(t)$ at different times cannot be related by unitary evolutions, not even short duration of time), but it is caused by the fact that the classical world with quantum correlation between nearby particles is contained within the universal wave function with diffusion wave packets all over the places. It seems plausible to assume that $\bar\rho_d(\vec x,t)/d$ is the mass density distribution calculated from $\Psi_u(t)$.

\section{Conclusions}
As a summary, we discover that the universal wave function can be far richer than Everett's many-world universal wave function. By analyzing the multi-body Schr\"{o}dinger equation and structure of the universal wave function, we found that the universal wave function can have special and observable gravitational effects, and that makes the universal wave function a realistic entity. Everett's formal solution is only the special model of our theory with $d=0$. In this case, $\Psi_c=\Psi_\chi=\Psi_u$ for a single classical world. {As a comparison, most studies based on Everett's idea are about the mechanism of classical world branching and decoherence based on the formal solution \cite{Everett} , rather than the gravitational effect.}

Here we wish to emphasize that even though we provided a plausible solution to explain the gravitational effects of dark matter, our model does not exclude the possibility of existence of new elementary particles \cite{Dark} which also create extra gravitational contributions, such as axion model \cite{axion1,axion2,axion3}. Based on quantum origin of the universe, we believe that the $\chi$ function presented in this article are inevitable consequences of evolutions of the universe. Once those large scale diffusion wave packets and their gravitational effects in the universal wave function are verified, the existence of many worlds becomes a universal truth.
Our researches provide new directions for future astronomical observation of the universe; that is, obtaining wave packets distribution information in the universal wave function through the observation of the gravitational effect of $\rho_d(\vec r,t)$. 

{Finally, we discuss several possible falsifiable predictions based on our theory.}

\begin{enumerate}
\item {In the application of our theory to dark matter, we do not consider deviations of gravitational force due to fluctuations of $\gamma_j$ and $\zeta_j$, and different $\gamma_j$ and $\zeta_j$ for different particles. Because the gravitational constant is very small, all observed gravitational effects are for systems consisting of a huge number of particles. In this case, it is extremely hard to notice differences in gravitational effects caused by fluctuations of $\gamma_j$ and $\zeta_j$. It would be interesting to notice that after many years of the earth-based measurements \cite{Luo} of Newtonian gravitational constant, there is an unexplained uncertainty in the measurement result of $G$  \cite{Luo}. One of the relevant experimental groups is asking for possible new physical principle \cite{Luo} to explain the uncertainty. Whether this kind of fluctuations of $\gamma_j$ and $\zeta_j$ has chances to interpret the uncertainty of Newtonian gravitational constant is a question worthwhile of researching in the future. 
In future work, we will consider dedicated experimental scheme to detect this fluctuating gravitational effect, which has the chance to reveal the gravitational force due to diffusion wave packets, aside from astronomical observations. If the unexplained uncertainty in the earth-based measurements \cite{Luo} of Newtonian gravitational constant can be found to have relevance to the fluctuation of $\gamma_j$ and $\zeta_j$, it would be a strong support to our theory.}

\item {In future work, we may consider the numerical calculations of Equation (\ref{simulation}) to simulate the large scale structure of our universe, so that to give a reliable test of our theory. Of course, this needs the adjustment of the initial condition of the universal wave function, and has the chance to provides further clue for the initial condition of the cosmic inflation.}

\item  {Our work not only shows that we need to change our understanding of gravitational effects in the classical world, but it also reveals that the Newtonian gravitational constant can be very different from the gravitational constant between elementary particles. Therefore, we also need to redefine magnitude of the gravitational constant between elementary particles, which provides the chance to test our theory in future. From the new gravitational constant $G=(d+1)G_N$, we should reconsider the grand unified theory and may provide indirect method to test our theory.}

\item {We may consider an indirect test of our theory by simulating the model with ultracold atomic gases \cite{AtomsGas}. First we prepare a completely isolated quantum gas in which all particle wave packets are diffusion wave packets. At the beginning we can even assume that there are no interactions between particles. Then, we artificially adjust the interaction energy to make phase transitions in this quantum gas. We use appropriate interaction energy to control the ratio of diffusion and localized parts after phase transitions took place. Under this condition, unitary evolutions can make the coexistence of different $\Psi_\chi(t)$ and $\Psi_c(t)$ with definite structures. With rapid developments of techniques for ultra-cold atoms gas such as Feshbach resonance \cite{AtomsGas,resonance} to control the interaction between atoms, we believe in the future there are prospects to simulate evolutions of the universe revealed in this article.}

%Finally, we point out the classical world wave function we presented here can have important implications in the field of condensed matter physics, for example maybe we can use correlated diffusion multi-body wave functions to analyze some incomprehensible properties of quantum material. Moreover, we believe those classical world wave functions can be simulated in the laboratory, here we take ultra cold atoms gas \cite{UltraCold} as an example to consider a simple experimental scheme.
\par
{When the temperature is extremely low, we may prepare the wave packets of ultra cold atoms gas with long range interactions contained within a potential \cite{AtomsGas} diffuse over a macroscopic distance. We can superpose a periodic potential on this system and position all the atoms at bottom of the periodic potential \cite{AtomsGas,lattice}. Now we can employ interactions such as Feshbach resonance \cite{AtomsGas,resonance} to form some correlated definite structures (such as artificial crystal). Under this condition, the crystal structure is correlated with the diffusion parts. The interaction energy for this system becomes $E=E_{c}+E_d+E_{cd}$, where $E_{cd}$ is the interaction energy between correlated crystal and diffusion part, $E_c$ is internal interaction energy for the crystal and $E_d$ stands for internal interaction energy of diffusion parts. Experimentally, people can measure the total interaction energy to determine if $E_{cd}$ really exists. Therefore this experimental scheme provides a way to simulate classical world wave functions presented in this article and the formation of classical world structures. Finally, even if we abandon motivations to simulate the universe, this problem per se is worthwhile of researching in the field of quantum multi-body manipulations.}
\end{enumerate}

{Note added: After the submission of our manuscript, the referee noticed a relevant paper \cite{Ernest} which argues that the solution of the Schr\"odinger equation including the gravitational potential may provide the possibility  to explain dark matter. However, the mechanism to explain dark matter is completely different from our manuscript. In \cite{Ernest}, the single-particle Schr\"odinger equation is used to explain the dark matter gravitational effect; while in our work the diffusion wave packet for the many-body universe wave function is shown to explain the gravitational effect of dark matter.}
%%%%%%%%%%%%%%%%%%%%%%%%%%%%%%%%%%%%%%%%%%
\vspace{6pt} 

%%%%%%%%%%%%%%%%%%%%%%%%%%%%%%%%%%%%%%%%%%
%% optional
%\supplementary{The following are available online at \linksupplementary{s1}, Figure S1: title, Table S1: title, Video S1: title.}


%%%%%%%%%%%%%%%%%%%%%%%%%%%%%%%%%%%%%%%%%%

\authorcontributions{conceptualization, {H.X.} and {Y.X.}; methodology, H.X. and Y.X.;  formal analysis, H.X. and Y.X.; investigation, H.X. and Y.X.;  writing---original draft preparation, H.X. and Y.X.; writing---review and editing, H.X. and Y.X. All authors have read and agreed to the published version of the manuscript.} %MDPI: we changed H.W. into H.X., changed Y.N. into Y.X., please check

%%%%%%%%%%%%%%%%%%%%%%%%%%%%%%%%%%%%%%%%%%
\funding{This research was funded by the National Natural Science Foundation of China under grants number 11334001, and 11175246.}

\institutionalreview{{Not applicable.}} %mdpi: {In this section, please add the Institutional Review Board Statement and approval number for studies involving humans or animals. Please note that the Editorial Office might ask you for further information. Please add “The study was conducted according to the guidelines of the Declaration of Helsinki, and approved by the Institutional Review Board (or Ethics Committee) of NAME OF INSTITUTE (protocol code XXX and date of approval).” OR “Ethical review and approval were waived for this study, due to REASON (please provide a detailed justification).” OR “Not applicable” for studies not involving humans or animals. You might also choose to ex-clude this statement if the study did not involve humans or animals.}

\informedconsent{{Not applicable.}} %mdpi:{Any research article describing a study involving humans should contain this statement. Please add “Informed consent was obtained from all subjects involved in the study.” OR “Patient con-sent was waived due to REASON (please provide a detailed justification).” OR “Not applicable” for studies not involving humans. You might also choose to exclude this statement if the study did not involve humans. Written informed consent for publication must be obtained from participating patients who can be identified (including by the patients themselves). Please state “Written informed consent has been obtained from the patient(s) to publish this paper” if applicable.}

\dataavailability{{Please refer to suggested Data Availability Statements in section "MDPI Research Data Policies" at https://www.mdpi.com/ethics.}} %mdpi:{In this section, please provide details regarding where data supporting reported results can be found, including links to publicly archived datasets analyzed or generated during the study. Please refer to suggested Data Availability Statements in section “MDPI Research Data Policies” at \href{https://www.mdpi.com/ethics}{https://www.mdpi.com/ethics}. You might choose to exclude this statement if the study did not report any data.} 

%%%%%%%%%%%%%%%%%%%%%%%%%%%%%%%%%%%%%%%%%%
\acknowledgments{We are grateful to X. S. Chen {and X. Q. Lin} for suggestions. }

%%%%%%%%%%%%%%%%%%%%%%%%%%%%%%%%%%%%%%%%%%
\conflictsofinterest{The authors declare no conflict of interest.} 

%%%%%%%%%%%%%%%%%%%%%%%%%%%%%%%%%%%%%%%%%%
%% Only for journal Encyclopedia
%\entrylink{The Link to this entry published on the encyclopedia platform.}

%%%%%%%%%%%%%%%%%%%%%%%%%%%%%%%%%%%%%%%%%%
%% Optional
%%%%%%%%%%%%%%%%%%%%%%%%%%%%%%%%%%%%%%%%%%
%% Optional
\appendixtitles{no} % Leave argument "no" if all appendix headings stay EMPTY (then no dot is printed after "Appendix A"). If the appendix sections contain a heading then change the argument to "yes".
\appendix
\setcounter{equation}{0}
\renewcommand\theequation{A.\arabic{equation}}
\renewcommand\thesection{A.\arabic{section}}

\setlength{\parindent}{2em}

%%%%%%%%%%%%%%%%%%%%%%%%%%%%%%%%%%%%%%%%%%
\end{paracol}
\reftitle{References}

\end{document}